\documentclass{article}
\usepackage{enumerate}
\usepackage{amsmath, amssymb, amscd}
\begin{document}


%
%

\title{A Variation of Levin Search for All Well-Defined Problems}

\author{Fouad B. Chedid\\A'Sharqiyah University, Ibra, Oman\\ f.chedid@asu.edu.om}

\maketitle


\begin{abstract}
In 1973, L.A. Levin published an algorithm that solves any inversion problem $\pi$ as quickly as the fastest algorithm $p^*$ computing a solution for $\pi$ in time bounded by $2^{l(p^*)}.t^*$, where $l(p^*)$ is the length of the binary encoding of $p^*$,  and $t^*$ is the runtime of $p^*$ plus the time to verify its correctness. In 2002, M. Hutter published an algorithm that solves any well-defined problem $\pi$ as quickly as the fastest algorithm $p^*$ computing a solution for $\pi$ in time bounded by $5.t_{p}(x)+d_p.time_{t_{p}}(x)+c_p$, where $d_p=40.2^{l(p)+l(t_{p})}$ and $c_p=40.2^{l(f)+1}.O(l(f)^2)$, where $l(f)$ is the length of the binary encoding of a proof $f$ that produces a pair $(p,t_p)$, where $t_p(x)$ is a provable time bound on the runtime of the fastest program $p$ provably equivalent to $p^*$. In this paper, we rewrite Levin Search using the ideas of Hutter so that we have a new simple algorithm that solves any well-defined problem $\pi$ as quickly as the fastest algorithm $p^*$ computing a solution for $\pi$ in time bounded by $O(l(f)^2).t_p(x)$. 
\end{abstract}
\vskip.1in\noindent
{\bf keywords:} Computational Complexity; Algorithmic Information Theory; Levin Search.

\section{Introduction}
We recall that the class $NP$ is the set of all decision problems that can be solved efficiently on a nondeterministic Turing Machine. Alternatively, $NP$ is the set of all decision problems whose guessed solutions can be verified efficienlty on a deterministic Turing Machine, or equivalently today's computers. The interesting part is that the set of all hardest problems in the class $NP$ (also known as $NP$-complete problems) have so far defied any efficient solutions on real computers. This has been a very frustrating challenge given that many of the $NP$-complete problems are about real-world applications that we really would like to be able to solve efficiently.
  
Leonid Levin \cite{levin}, and independently from Stephen Cook \cite{cook}, defined the class $NP$, identified the class $NP$-complete, and gave a few reductions that show the $NP$-completeness of some $NP$ problems (independent from the work of Richard Karp \cite{karp}).  In that same paper \cite{levin}, Levin gave an algorithm to deal with $NP$-complete problems provided that someone develops a proof that shows $P=NP$. In this case, Levin's algorithm can be used to develop a polynomial-time solution for every $NP$-complete problem. In particular, Levin's algorithm works with a very broad class of mathematical problems that can be put in the form of inverting easily computable functions. For example, suppose we have a function $y=f(x)$, where we know nothing about $f$ except that it is easily computable. The challenge is to find $x=f^{-1}(y)$ and to do so in a minimal amount of time. The idea of Levin's algorithm is to go through the space of all algorithms in search for a fastest algorithm $A$ that knows the secret of $f$. In general, let $\pi$ be an inversion problem and let $p^*$ be a known fastest algorithm for $\pi$ that runs in time $t^*$, where $t^*$ is the runtime of $p^*$ plus the time to verify its correctness. The Universal Search algorithm of Levin, also known as Levin Search, is an effective procedure for findng $p^*$ in time bounded by $2^{l(p^*)}.t^*$, where $l(p^*)$ is the length of the binary encoding of $p^*$. Levin Search can also be used with some forms of optimization and prediction problems, and it is theoretically optimal. The following is a pseudocode for Levin Search:
\vskip.1in\noindent
{\bf Pseudocode Levin Search}\\
$t = 2$.\\
for all programs $p$, in parallel, do\\
\indent run $p$ for at most $t.2^{-l(p)}$ steps\\
\indent if $p$ is proved to generate a correct solution for $\pi$, then return $p$ for $p^*$ and halt.\\
endfor\\
$t = 2.t$\\
goto for all programs $p$\\
{\bf End of Pseudocode}\\
\vskip.05in\noindent
The idea of Levin Search is to search the space of all programs $p$ in increasing order of $l(p)+\log t$, or equivalently in decreasing order of $\frac {Prob.(p)} t$, where $Prob.(p)=2^{-l(p)}$ is the probability contributed to $p$ in the overall Solomonoff's algorithmic probability \cite{ray64} of finding a solution for $\pi$. By construction, $Prob.(p)$ gives short programs a better chance of being successful on the assumption that short programs are more worthy (Occam's razor). We mention that the main idea of Levin Search was behind the notion of Levin Complexity, which defines a computable version of Kolmogorov complexity based on time-bounded Turing machines. Let $U$ be a fixed reference universal Turing machine, the Levin complexity of a finite binary string $x$ is defined as 
$$K_t(x)=\min._{p}\{l(p)+\log t: U(p) = x \text{, in at most }t \text{ steps.}\}$$   
Returning to Levin Search, during each iteration of the ``for'' loop, each program is assigned a fraction of time proportional to its probability, which basically says that short programs get to run more often.

The rest of this paper is organized as follows. Section 2 describes a sequential implementation of Levin Search and shows that it solves any inversion problem $\pi$ as quickly as the fastest algorithm $p^*$ computing a solution for $\pi$, save for a factor of $2^{l(p^*)}$. Section 3 describes an improvement on Levin Search for all well-defined problems.  A problem is designated as well-defined if its solutions can be formally proved correct and have provably quickly computable time bounds. This improvement, known as Hutter's algorithm \cite{hutter}, describes an algorithm that solves any well-defined problem $\pi$ as quickly as the fastest algorithm computing a solution for $\pi$, save for a factor of $5$. Section 4 contains our contribution. We rewrite Levin Search using the ideas of Hutter so that we have a new simple algorithm that solves any well-defined problem $\pi$ as quickly as the fastest algorithm $p^*$ computing a solution for $\pi$, save for a factor of  $O(l(f)^2)$, where $l(f)$ is the length of the binary encoding of a proof $f$ that produces a pair $(p,t_p)$, where $t_p$ is a provable time bound on the runtime of the fastest program $p$ provably equivalent to $p^*$. Section 5 contains some concluding remarks.
\section{Sequential Levin Search} 
The pseudocode of Levin Search from the previous section cannot be readily executed on today's computers. A sequential version of Levin Search can simulate the parallel execution of all programs by running programs, each for a fraction of time, one after another, in increasing order of their length. Hereafter, we assume that all progams are encoded using prefix-free codes, using for example, Shannon-Fano coding \cite{shannon}. Let $\pi$ be an inversion problem and let $p^*$ be a known fastest algorithm for $\pi$ that runs in time $t^*$, where $t^*$ is the runtime of $p^*$ plus the time to verify its correctness. Also, let $l(p)$ be the length of the binary encoding of program $p$.
\vskip.1in\noindent
{\bf Algorithm Levin Search}\\
$t = 2$;\\
for all programs $p$ of length $l(p)=1,2,\ldots,\log t$ do \\
\indent run $p$ for at most $t.2^{-l(p)}$ steps.  \\
\indent if $p$ is proved to generate a correct solution for $p$, return $p$ for $p^*$  and halt.\\
endfor\\
$t = 2.t$\\
goto for all programs $p$\\
{\bf End of Levin Search}
\vskip.1in\noindent
During each time step $t$, the algorithm runs all programs $p$ of length $\le \log t$ in increasing order of their length and gives each program $p$ a fraction of time proportional to $2^{-l(p)}$.
\vskip.1in\noindent
\noindent{\bf Time Analysis.} The following proof for the runtime complexity of Levin Search is due to Solomonoff \cite{ray85}. The ``for'' loop in the algorithm takes time $\sum\limits_{p: l(p)\le \log t} t.2^{-l(p)}\le t$. This is true by Kraft inequality \cite{kraft}. Now, suppose that Levin Search stops at time $t = T$. Then, the total runtime of this algorithm is 
$$time_{Levin Search} \le 2+4+\ldots +\frac T 4+\frac T 2+T= 2T-2<2T.$$
Moreover, the runtime $t^*$ of $p^*$ is $$t^*= T.2^{-l(p^*)}>\frac {time_{Levin Search}} 2.2^{-l(p^*).}$$
Thus, $$time_{Levin Search} < 2^{l(p^*)+1}.t^*=O(t^*)$$
This is true because the multiplicative constant $ 2^{l(p^*)+1}$ is independent of the instance of the problem $\pi$. We mention that it is this huge multiplicative constant that limits the applicability of Levin Search in practice.

It seems that a possible way to improve the runtime of Levin Search is to find a way, where not all programs of length $\le \log t$ get executed, for each value of $t$. Ideally, we would like a way that executes at most a single program for each value of $t$, preferably a shortest and fastest one, and this is exactly what we will present in Section 4.

\section{Hutter's Fastest and Shortest Algorithm for All Well-Defined Problems}
In \cite{hutter}, Hutter presented an improvement on Levin Search for problems that are well-defined. By this, Hutter means a problem whose solutions can be formally proved correct and have provably quickly computable time bounds. Let $p^*(x)$ be a known fastest algorithm for some well-defined problem $\pi(x)$. Then, Hutter's algorithm will construct a solution $p(x)$ that is provably equivalent to $p^*(x)$, for all $x$, in time proportional to $5.t_{p}(x)+d.time_{t_{p}}(x)+c$, where $t_{p}(x)$ is a provable time bound on the runtime of $p(x)$, $time_{t_{p}}(x)$ is the time needed to compute $t_{p}(x)$, and $d$ and $c$ are constants which depend on $p$ but not on $\pi(x)$. This shows that Hutter's agorithm runs in time $O(t_{p})$, save for a factor of 5. 

Our work in the next section is inspired by the ideas of Hutter, and so, for completeness, we have included below the details of Hutter's algorithm as they appear in \cite{hutter}. 

The main idea of Hutter's algorithm is to search the space of all proofs\footnote{Given a formal logic system $F$ with a set of axioms and inference rules. A proof in $F$ is a sequence of formulas, where each formula is either an axiom in $F$ or is something that can be inferred using $F$'s axioms and inference rules, and previoulsy inferred formulas.} in some formal axiomatic system, and not the space of all programs. In particular, the algorithm searches for those proofs that can tell us which programs are provably equivalent to $p^*$ and have provably quickly computable time bounds. This is doable since the set of all proofs in any formal sysem is enumerable.  Moreover, in \cite{hutter}, Hutter showed how to formalize the notions of provability, Turing Machines, and computation time. Let $U$ be a fixed reference Universal Turing machine. For finite binary strings $p$ and $t_p$, we assume that 
\begin{itemize}
\item $U(p,x)$ computes the output of program $p$ on input $x$.
\item $U(t_p,x)$ returns a time bound on the runtime of program $p$ on input $x$.  
\end{itemize}
Moreover, Hutter defined the following two terms:
\begin{itemize}
\item A term $u$ is defined such that the formula $[\forall y: u(p,y)=u(p^*,y)]$ is true if and only if $U(p,x)=U(p^*,x), \forall x$.
\item A term $tm$ is defined such that the formula $[tm(p,x)=n]$ is true if and only if $U(p,x)$ takes $n$ steps; that is, if $time_p(x)=n$.
\end{itemize}
Then, we say that programs $p$ and $p^*$ are provably equivalent if the formula  $[\forall y: u(p,y)=u(p^*,y)]$ can be proved.
\vskip.1in\noindent
The algorithm of Hutter is as follows.
\vskip.1in\noindent
{\bf Algorithm Hutter($x$)}\\
Let $L=\phi$, $t_{fast}=\infty$, and $p_{fast}=p^*$.\\
Run A,B, and C concurrently with 10\%, 10\%, and 80\% of computational resources, respectively.
\vskip.1in\noindent
{\bf Algorithm A}\\
\{ This algorithm identifies programs $p$ that are provably equivalent to $p^*$ and have provably computable time bounds $t_p$ \}\\
for $i=1,2,3,\ldots$ do \\
\indent pick the $i^{th}$ proof in the list of all proofs and\\
\indent if the last formula in the proof is equal to $[\forall y, u(p^*,x)=u(p,y)$ and \\
\indent $u(t_p,y) \ge tm(p,y)]$, for some pair of strings $(p,t_p)$ then \\
\indent\indent Let $L=L\cup \{(p,t_p)\}$.\\
{\bf End A}
\vskip.1in\noindent
{\bf Algorithm B}\\
\{ This algorithm finds the program with the shortest time bound in $L$.\}    \\
for all $(p,t_p)$ in $L$ do\\
\indent run $U$ on all $(t_p,x)$ in parallel for all $t_p$ with relative computational \\
\indent resources $2^{-l(p)-l(t_p)}$.\\
\indent if $U$ halts for some $t_p$ and $U(t_p,x)< t_{fast}$ then\\
\indent\indent $t_{fast} = U(t_p,x)$ and $p_{fast}=p$.\\ 
endfor all $(p,t_p)$\\
{\bf End B}
\vskip.1in\noindent
{\bf Algorithm C}\\
For $k=1,2,4,8,\ldots$ do\\
\indent pick the currenlty fastest program $p_{fast}$ with time bound $t_{fast}$ .\\
\indent run $U$ on $(p,x)$ for $k$ steps.\\
\indent if $U$ halts in less than $k$ steps, then print result $U(p,x)$ and abort computation \\
\indent of A, B, and C.\\
endfor $k$\\
{\bf End C}
\vskip.1in\noindent
In \cite{hutter}, it was shown that the overall runtime of Hutter's algorithm is $$time_{Hutter}(x)\le 5.t_{p_{fast}}(x)+d_p.time_{t_{p_{fast}}}(x)+c_p,$$ 
where $d_p=40.2^{l(p_{fast})+l(t_{p_{fast}})}$ and $c_p=40.2^{l(proof(p_{fast}))+1}.O(l(proof(p_{fast})^2)$.
\vskip.1in\noindent
\section{A Variation of Levin Search Inspired by the Fastest and Shortest Algorithm of Hutter}
We rewrite Levin Search so that it searches, in parallel, the space of all proofs, and not the space of all programs. Let $p^*$ be a known fastest algorithm for some well-defined problem $\pi$. Then, our algorithm will construct a solution $p(x)$, for all $x$, that is provably equivalent to $p^*(x)$ in time $O(t_p(x))$, save for a factor of $O(l(proof(p))^2)$, where $t_{p}(x)$ is a provable time bound on the runtime of the fastest program $p$ provably equivalent to $p^*$.
\vskip.1in\noindent 
We assume a formal axiomatic system $\mathbb F$ and the same terms $u$ and $tm$ defined in Hutter's algorithm.  Moreover, we assume that all pairs $(p,t_p)$ are encoded using prefix-free codes, and hence, we have $\sum\limits_{(p,t_p)} 2^{-l(p)-l(t_p)}\le 1$, by Kraft inequality.
\vskip.1in\noindent
The following is our Modified Levin Search algorithm:
\vskip.1in\noindent
{\bf Algorithm Modified Levin Search($x$)}\\
$t = 2$; $t_{fast}=\infty$; $p_{fast}=\lambda$; \{ $p_{fast}$ is initialized to the empty string \} \\
for all proofs $f\in F$ of length $l(f)=1,2,\ldots,\log t$ do \\
\indent write down the first $t.2^{-l(f)}$ characters of proof $f$.\\
\indent if the last formula in these characters is equal to $\forall y, u(p^*,x)=u(p,y)$ \\
\indent and $u(t_p,y) \ge tm(p,y)$, for some pair of strings $(p,t_p)$ then \\
\indent\indent run $t_p(x)$ for at most $t.2^{-l(p)-l(t_p)}$ steps.\\
\indent\indent if $t_p$ halts and $t_p(x) < t_{fast}$ then\\
\indent\indent\indent $t_{fast} = t_p(x)$ and $p_{fast}=p$.\\ 
endfor\\
if $p_{fast} \not= \lambda$ then\\
\indent run $p_{fast}$ for at most $t$ steps with time bound $t_{fast}$. \\
\indent if $p_{fast}(x)$ halts, then return $p_{fast}$ for $p^*$ and halt. \\
endif \\
$t = 2t.$\\
go to for all proofs $f\in F$\\
{\bf End of Modified Levin Search}
\vskip.1in\noindent
We note that during each time step $t$, Modified Levin Search runs at most a single provably fast program $p_{fast}$ that is provably equivalent to $p^*$. 
\vskip.1in\noindent
{\bf Running time calculation: } Following \cite{hutter}, let $n_{axioms}$ be the finite number of axioms in $\mathbb F$. Then, each proof $f\in \mathbb F$ is a sequence $F_1F_2\ldots F_n$, for some positive integer $n$, where each $F_i$ $(1\le i\le n)$ is either an axiom or a formula in $\mathbb F$. It takes $O(n_{axioms}.l(F_i))$ time to check if $F_i$ is an axiom, and $O(l(f))$ time to check if $F_i$ is a formula that can be inferred from other formulas $F_j\in f$, $j<i$. Thus, checking the validity of all formulas in $f$, and hence the validity of the proof $f$, takes $O(l(f)^2)$ time. 
\vskip.1in\noindent
The ``for'' loop in Modified Levin Search takes time 
\begin{align*}
t_1&=\sum\limits_{f: l(f)\le \log t} O(t.2^{-l(f)})+O(( t.2^{-l(f)})^2) + t.2^{-l(p)-l(t_p)}\\
&\le t+\sum\limits_{f: l(f)\le \log t} O(l(f))+O(l(f)^2)  
\end{align*}
This is true because  $\sum\limits_{(p,t_p)} 2^{-l(p)-l(t_p)}\le 1$, by Kraft inequality. we next have
\begin{align*}
t_1&\le t+\sum\limits_{f: l(f)\le \log t}O(l(f)^2)\\ 
&< t + 2^{\log t+1}.O(l(f^{\prime})^2)\\
&< t + 2t.O(l(f^{\prime})^2)\\
&< O(l(f^{\prime})^2).t,
\end{align*}
where $f^{\prime}$ is the proof that procuded the pair $(p,t_p)$ with the shortest provable time bound $t_p(x)$, among all proofs of length $\le \log t$. 
\vskip.1in\noindent
Thus, the runtime of Modified Levin Search is equal to
\begin{align*}
time_{Modified Levin Search} &=\sum_{t=2} O(l(f^{\prime})^2).t + t \\
&=\sum_{t=2} O(l(f^{\prime})^2).t\\
&<\sum_{t=2} O(l(f^*)^2).t,
\end{align*}
where $l(f^*)$ is the length of the binary encoding of the proof that produced the resultant pair $(p_{fast},t_{p_{fast}})$.
\vskip.1in\noindent
Suppose that Modified Levin Search stops at time $t=T\le t_{p_{fast}}(x)$.Then
\begin{align*}
time_{Modified Levin Search}(x) &= O(l(f^*)^2)\}.(2+4+\ldots+T/2+T) \\
&=O(l(f^*)^2).(2T-2) \\
&< O(l(f^*)^2).2T \\
&=O(l(f^*)^2).T \\
&\le O(l(f^*)^2).t_{p_{fast}}(x).
\end{align*}
This shows that Modified Levin Search runs in time $O(t_{p_{fast}})$, save for a factor of $O(l(f^*)^2)$.

\section{Concluding Remarks}
We recall that Hutter's algorithm was designed for well-defined problems. These are problems the solutions of which can be formally proved correct and have provably quickly computable time bounds. For programs provably correct (they halt and compute the correct answer), but for which no quickly computable time bounds exist (For example, the traveling salesman problem), Hutter \cite{hutter} explained that an obvious time bound for a progam $p$ is its actual running time $time_p(.)$. By replacing $t_p(.)$ with $time_p(.)$ in the runtime of Hutter's algorithm, and by noticing that, in this case, $time_{t_p}(x)$ becomes $time_{time_p}(x) \le time_p(x)$, we have 
\begin{align*}
time_{Hutter}(x) &\le 5.t_{p}(x)+d_p.time_{t_{p}}(x)+c_p \\
&\le 5.time_p(x)+d_p.time_p(x)+c_p\\
&\le (5+d_p).time_p(x) + c_p\\
&\le (5+40.2^{l(p)}).time_p(x)+40.2^{l(f)+1}.O(l(f)^2)
\end{align*}
where $l(p)$ is the length of the binary encoding of $p$, and $l(f)$ is the length of the binary encoding of a proof $f$ that produces the program $p$ that is provably equivalent to the fastest known algorithm $p^*$ for the problem in hand. Thus, for such programs, Hutter's algorithm is optimal, save for a huge constant multiplicative term and a huge constant additive term.

We next calculate the runtime of our algorithm in case we are dealing with programs that can be provably correct, but for which no quickly computable time bounds exist. Suppose that our algorithm stops at time $t=T$. Then, we have 
\begin{align*}
time_{Modified Levin Search} &<O(l(f)^2).T.
\end{align*}
We also have $time_{p}(x)=T.2^{-l(p)}$. Thus,
$$time_{Modified Levin Search} < O(l(f)^2).{2^{l(p)}.time_{p}(x)} $$
Thus, for such programs, the runtime of our algorithm also suffers from a huge constant multiplicative term, but with no additinoal huge additive terms.

We conclude this paper by recalling an approach, proposed by Solomonoff in 1985 \cite{ray85}, to manage the huge multiplicative constant in the big-oh notation of the running time of Levin Search. We do so because we think that the importance of that work of Solomonoff is not widely appreciated. Solomonoff argued that if machines are going to have a problem solving capability similar to that of humans, then machines cannot start from scratch everytime they attempt to solve a new problem. We, humans, rely on our previous knolwedge of solutions to other problems to figure out a solution for a new problem. The basic idea of Solomonoff is that we should be able to construct $p^*$ via a list of references to previously discovered solutions for other related problems. We can imagine writing the code for $p^*$ as $p_1p_2\ldots p_n$, where $p_i$ is a reference to a solution for problem $\pi_i$ stored on a separate work tape of a Kolmogorov-Uspensky Machine \cite{kol}. This way, the solution $p^*$ is a sequence of calls to other solutions stored on the work tapes. This way, the length of $p^*$ would be made significantly smaller than the sum of the lengths of $p_i$, and the saving in the length of $p^*$ would exponentially decrease the multiplicate constant in the big-oh notation of Levin Search.

\end{document}